\documentclass[twocolumn,showpacs,amssymb,nobibnotes,aps,floats,psfig,prl]{revtex4}
\usepackage{hyperref}
\usepackage{amsbsy}
\usepackage{amsmath}
\usepackage{graphicx,epsfig}
\usepackage{epsfig}
\usepackage{float}
\def \be{\begin{equation}} 
\def \ee{\end{equation}} 
\def \bea{\begin{eqnarray}} 
\def \eea{\end{eqnarray}}
\begin{document}
\title{Supersolidity in a Bose-Holstein model}
\author{Sanjoy Datta  }
%\email{sanjoy.datta@saha.ac.in}
\author{Sudhakar Yarlagadda}
%\email{y.sudhakar@saha.ac.in}
\affiliation{CAMCS and TCMP Div., Saha Institute of Nuclear Physics, 
1/AF Salt Lake, Kolkata-64, India}
\date{\today}

\begin{abstract}
We derive an effective d-dimensional Hamiltonian for a system of 
hard-core-bosons coupled to optical phonons in a lattice. At non-half-fillings, 
a superfluid-supersolid transition occurs at intermediate boson-phonon
couplings, while at strong-couplings the system phase separates. We demonstrate
explicitly that the presence of next-nearest-neighbor hopping and 
nearest-neighbor repulsion leads to supersolidity.
Thus we present a microscopic mechanism for the
homogeneous coexistence of
charge-density-wave and superfluid orders.
\end{abstract}
\pacs{67.80.kb,  64.70.Tg,  71.38.-k,  71.45.Lr }
%***********************************************************************
\maketitle

\nopagebreak
{\bf Introduction:} 
Coexistence of charge-density-wave (CDW)   
and superconductivity 
is manifested in a variety
of systems such as the perovskite type bismuthates (i.e.,
${\rm BaBiO_3}$ doped with ${\rm K}$ or ${\rm Pb}$) \cite{blanton},
quasi-two-dimensional layered dichalcogenides (e.g., 
${\rm 2H-TaSe_2}$, ${\rm 2H-TaS_2}$, and ${\rm 2H-NbSe_2}$)
 \cite{withers}, etc.
What is special about 
the above systems
is that they defy the usual expectation that the competing CDW
and superconductivity orders occur mutually exclusively.
CDW and superconductivity are examples of 
diagonal long range order (DLRO) and
off-diagonal long range order (ODLRO) respectively.
Here, DLRO breaks a continuous translational invariance  into a discrete
translational symmetry, whereas ODLRO breaks a global U(1)
phase rotational invariance \cite{penrose}.

 Another interesting example of DLRO-ODLRO concurrence is
a supersolid (SS). A SS state 
is characterized by the homogeneous
coexistence of two seemingly mutually contradictory phases, namely,
a crystalline solid and a superfluid (SF).
Here all particles simultaneously participate in both
 types of long range order.
It has been conjectured long ago that solid helium-4 may exhibit supersolidity 
due to Bose condensation of vacancies present
 \cite{Andreev,leggett}.
 Only recently, Chan and Kim 
(using a  torsional oscillator) observed a decoupling of
a small percentage of the helium solid from a container's walls
\cite{KimChan-Nature}.  They interpreted this as supersolidity.
 This discovery led 
to studies of bosonic models 
 in different kinds of
 lattice structures \cite{Damle} and with various 
types of particle interactions
\cite{Sengupta}.
Phenomenologically,
 Ginzburg-Landau theory \cite{jinwu} and a
quantum solid based on  Gross-Pitaevskii equation \cite{josserand}  have
been used to study supersolidity.

In this paper, we address the above ongoing puzzles by
studying the quantum phase transitions
exhibited by hard-core-bosons (HCB) coupled to optical phonons.
To this end, in contrast to the above treatments,
we employ a microscopic approach involving
 a minimal Bose-Holstein (BH) lattice model.
 Examples of real systems describable by our BH model are as follows.
 In the bismuthate systems, the observed valence skipping of the bismuth ion
is explained by invoking non-linear screening
which is said to produce a large attractive interaction
resulting in the formation of local pairs or HCB
\cite{varma,tvr}. Such HCB couple to the cooperative breathing mode
of the oxygen octahedra surrounding the Bismuth ion.
In a helium-4 crystal, vacancies produce a local distortion
and can be treated as HCB coupled to finite frequency phonons.
Furthermore, the concentration of vacancies is very small \cite{leggett}
and hence direct interactions among these HCB is negligible.
Lastly, for dichalcogenides such as ${\rm NbSe_2}$,
 where homogeneous coexistence
of the two long range orders has been unambiguously established
\cite{suderow}, our results
on the BH model should be quite relevant. 

\begin{figure}[b]
\includegraphics[angle=0,width=0.48\linewidth]{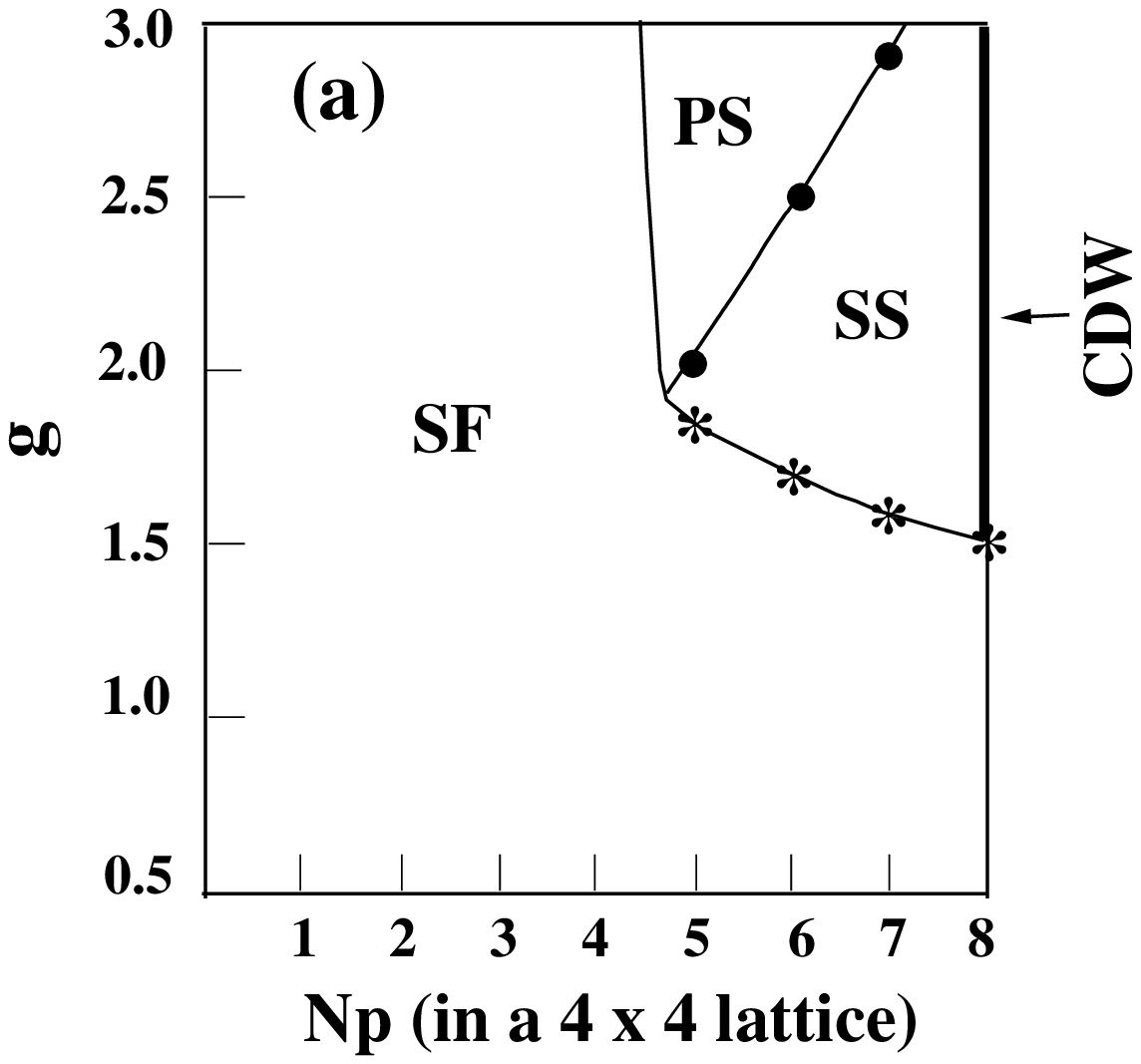}
\hspace{0.007\linewidth}\includegraphics[angle=0,width=0.48\linewidth]
{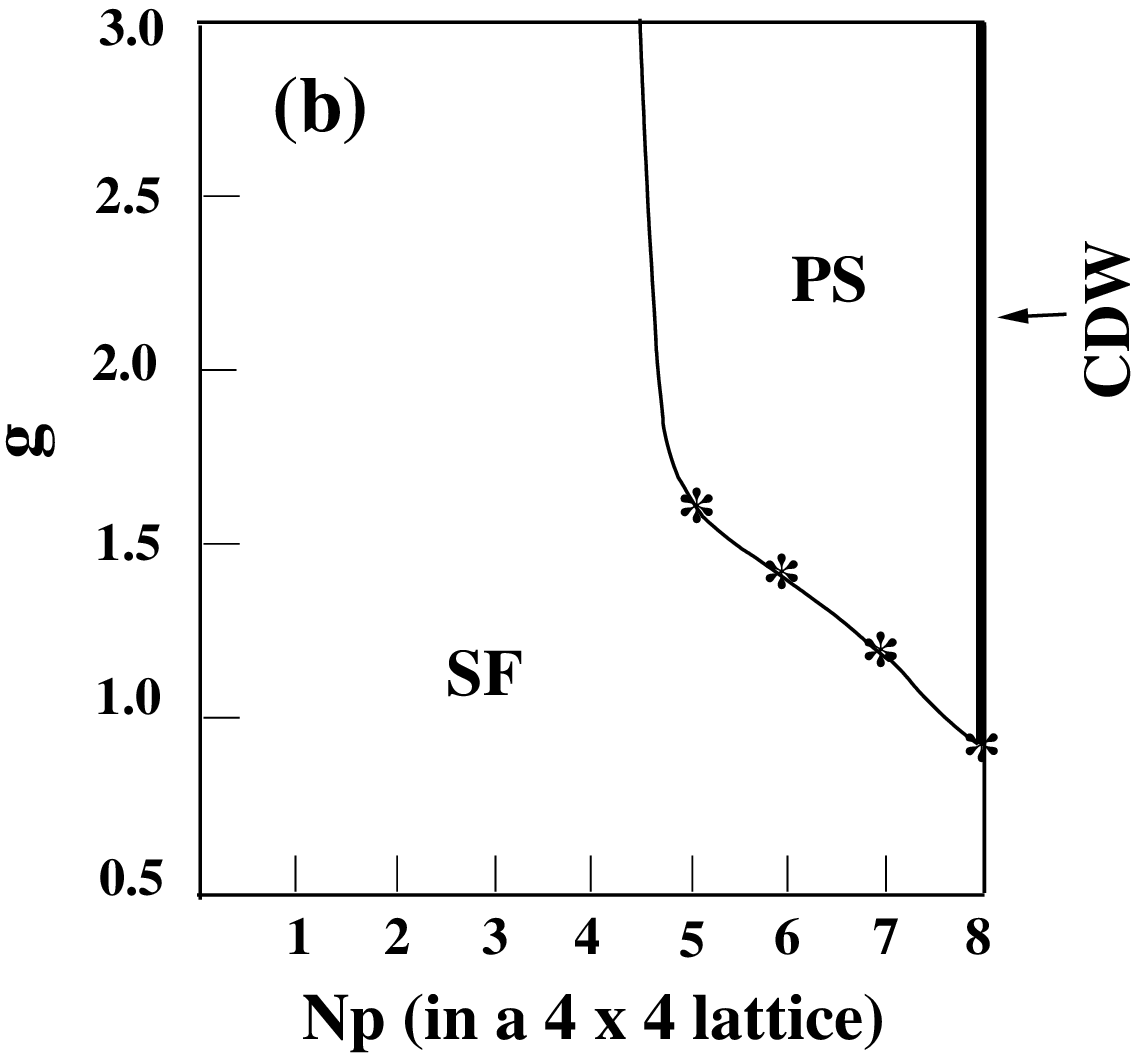}
\caption{Quantum phase diagram at various particle numbers $N_p$,
 $t/\omega_0 = 1.0$
and (a) NNN hopping $J_2 > 0$ in our effective
BH Hamiltonian [Eq. (\ref{heff})] and (b) $J_2 = 0$ 
(xxz-model). Supersolidity occurs only in (a).
The continuous lines in (a) and (b) are guides for the eye.
 }
\label{phase-diag2}
\end{figure} 
 Starting with our
BH model, we derive
an effective d-dimensional Hamiltonian for HCB 
by using a transparent non-perturbative technique.
The region of validity of our effective Hamiltonian is governed by the
small parameter ratio of the adiabaticity $t/\omega_0$ and the
boson-phonon (b-p) coupling $g$. The most interesting feature of this 
effective Hamiltonian is that
it contains an additional next-nearest-neighbor (NNN)
hopping compared to the Heisenberg xxz-model involving
only nearest-neighbor (NN) hopping and NN repulsion \cite{micnas,ranninger2}. 
We employed
a modified Lanczos algorithm \cite{gagli}
(on lattices of sizes $4 \times 4$, 
$\sqrt{18} \times \sqrt{18}$,  
$\sqrt{20} \times \sqrt{20}$,  
and $4 \times 6$) and found
that (except for the extreme anti-adiabatic limit)
{\em the BH model shows supersolidity at
intermediate b-p coupling strengths} 
whereas the xxz-model produces only a phase separated (PS) state.
Here we present the calculations for only a $4 \times 4$ lattice
(see Fig. \ref{phase-diag2}).
Our main results are depicted in 
Fig. \ref{phase-diag2} (a), where supersolidity manifested by
local pairs
(as HCB) implies homogeneous coexistence
of CDW and superconductivity.

{\bf Effective Bose-Holstein Hamiltonian:}
We start with a system of spinless HCB coupled with 
optical phonons on a square lattice. This system is described by a 
BH Hamiltonian 
\cite{ranninger1}
\be
H = -t \sum_{j,\delta}b^{\dagger}_j  b_{j+\delta} 
          + \omega_0 \sum_j a^{\dagger}_{j} a_j
          + g \omega_0 \sum_j n_j
 (a_j +a^{\dagger}_j) , 
\ee
where $\delta$ corresponds to nearest-neighbors,
$\omega_0$ is the optical phonon frequency,
 $b_j$ ($a_j$) is the destruction operator for HCB (phonons), and
 $n_j \equiv b^{\dagger}_{j} b_j $.
 Then we perform 
the Lang-Firsov (LF) transformation \cite {lang,sdadys} on this
Hamiltonian and this produces displaced
simple harmonic oscillators and dresses the hopping particles with phonons.
Under the LF transformation  given by $e^S H e^{-S} =H_0+H' $ with
$S= - g \sum_i n_i (a_i - a^{\dagger}_i)$,
$b_j$ and $a_j$ transform like fermions  and phonons in the Holstein model.
This is due to the unique (anti-) commutation properties of HCB given by 
\bea
[b_i,b_j]&=&[b_i,b^{\dagger}_j]= 0 , \textrm{ for } i \neq j , \nonumber\\
\{b_i,b^{\dagger}_i\}& = & 1 .
\label{commute}
\eea     
Next, we take the unperturbed Hamiltonian $H_0$ to be 
\cite{sdadys}
\be
H_0 = -J_1 \sum_{j,\delta} b^{\dagger}_j b_{j+\delta} +
 \omega_0 \sum_j a^{\dagger}_j a_j 
      - g^2 \omega_0 \sum_j
n_j ,
\ee 
and the perturbation $H'$ to be
\be
H'
 = -J_1 \sum_{j, \delta} b^{\dagger}_j b_{j+\delta}
            \{\mathcal S^{{j}^\dagger}_+ \mathcal S^{j}_{-}-1\} ,
\ee
where $\mathcal S^{j}_{\pm} = \textrm{exp}[\pm g(a_j - a_{j+\delta})],
J_1 = t \textrm{exp}(-g^2) $, and $g^2 \omega_0 $
is the polaronic binding energy. We then follow the same 
steps as in Ref. [\onlinecite{sdadys}] to get the following effective 
Hamiltonian in d-dimensions for our BH model
\bea
\!\!\!\!
H_{e}
\!\! &=&
\!\! -g^2 \omega_0 \sum_j n_j - 
         J_1 \sum_{j,\delta} b^{\dagger}_j b_{j+\delta} \nonumber \\
      & &
\!\!\!\!\!\!
 - J_2 \sum_{j,\delta,\delta' \neq \delta}
             b^{\dagger}_{j+\delta'}b_{j+\delta} 
         - 0.5 J_z \sum_{j,\delta} n_j(1- n_{j+\delta}) ,
\label{heff}
\eea  
where $J_z \equiv (J_1^2/\omega_0)[4 f_1(g)+2 f_2(g)]$ and 
$J_2 \equiv (J_1^2/\omega_0)f_1(g)$ with 
$f_1(g) \equiv \sum^{\infty}_{n=1} g^{2n}/(n!n)$ and
$f_2(g) \equiv \sum^{\infty}_{n=1}\sum^{\infty}_{m=1} g^{2(n+m)}/[n!m!(n+m)].$
Here we would like to point out that, as shown in Ref. \onlinecite{sudhakar1},
the small parameter for our perturbation theory is $t/(g \omega_0)$.

{\bf  Long range orders:}
Diagonal long range order (DLRO) can be characterized
by the structure factor defined
in terms of the particle density operators as follows:
\begin{equation}
S(\mathbf{q}) = \frac{1}{N}
\sum_{i,j} e^{\mathbf q \cdot(\mathbf R_i -\mathbf R_j )}
(\left\langle n_i n_j\right\rangle -
\left\langle n_i\right\rangle \left\langle n_j\right\rangle ).
\label{sf}
\end{equation}

Off-diagonal long range order (ODLRO) in a Bose-Einstein condensate (BEC), as
 introduced by Penrose and Onsager \cite{penrose}, 
is characterized by the order parameter 
$\left< b_0 \right> = \sqrt{\left<n_0 \right>} e^{i\theta}$ 
where $n_0$ is the occupation number for the $\mathbf k = 0$ momentum state or
the BEC.
 It is useful to define the general one-particle density matrix  
\begin{equation}
\tilde \rho (i,j) = \left< b^{\dagger}_i b_j \right> = 
\frac{1}{N} \sum_{\mathbf k,\mathbf q} e^{(\mathbf k \cdot 
\mathbf R_i - \mathbf q \cdot \mathbf R_j)}
\left< b^{\dagger}_{\mathbf k} b_{\mathbf q} \right> ,
\label{density_matrix}
\end{equation}
where $\left< \right>$ denotes ensemble average.
Eq. (\ref{density_matrix})
 gives the BEC fraction as 
\begin{equation}
n_b=
\frac{ \left<n_0\right>}{N_p}=
\sum_{i,j} \frac{\tilde \rho (i,j)}{N N_p} .
\label{n_b}
\end{equation}
In general, to find $n_b$, one
constructs the generalized one-particle density matrix $\tilde \rho$ 
and then diagonalizes it to find out
the largest eigenvalue. 
To characterize a SF, an important quantity is the 
 SF fraction $n_s$  which is calculated as follows.
Spatial variation in the phase 
 of the SF order parameter will increase
the free energy of the system.
We  
use a linear phase variation $\theta(x) = \theta_0 (x/L)$ 
with $\theta_0 $ being a small angle and $L$ the linear dimension in
x-direction.
This is done by imposing twisted boundary conditions (TBC)
 on the 
many-particle
 wave function.
At $T=0$ K, we can write the change in energy to be 
\begin{equation}
 E[\theta_0] - E[0]=
\frac{1}{2} m N_p n_s \left|\frac{\hbar}{m} \vec\nabla \theta(x) \right|^2 .
\end{equation}
Then the SF fraction is given by \cite{fisher,roth}
\begin{equation}
n_s = \left(\frac{N}{N_p t_{eff}}\right)
\frac{E[\theta_0] - E[0]}{\theta_0^2},
\label{sffrac_ours} 
\end{equation}
where $t_{eff} = \hbar^2/2 m$. For our Hamiltonian in Eq. (\ref{heff}),
we find $t_{eff} = J_1 + 8 J_2$.
The phase variation (taken to be the same for BEC and SF order parameters)
is introduced in our calculations
 by modifying the hopping terms
with $b_j  \rightarrow b_j \exp[i \hat{ \mathbf x}\cdot \mathbf R_{j}
 \theta_0/L ]$ which
is gauge-equivalent to TBC.

{\bf Results and discussion:}
We employ the mean field analysis (MFA) of Robaszkiewicz
 {\em et al.} \cite{micnas} to study the phase transitions
dictated by the effective Hamiltonian of Eq. (\ref{heff}). We
obtain the following expression for the SF-PS (SF-CDW) phase boundary 
at non-half-filling (half-filling): 
\bea
 \frac{J_{z}}{2 J_{1}} - \frac{3 J_{2}}{J_{1}} =
\frac{1+\left(2 n-1\right)^2}{1-\left(2 n-1\right)^2}  .
\label{phaseboundary}
\eea  
Eq.(\ref{phaseboundary})  leads to the same phase diagram
as that (for the xxz-model) in Ref. \onlinecite{micnas} but with 
$J^{eff}_{z} = J_z/2 - 3 J_2$ as the y-coordinate instead of
$J^{eff}_{z} = J_z/2 $.
Thus, within mean field, NNN hopping does not change the qualitative
features of the phase diagram; it
only increases the critical value of ${J^{}_{z}}/{J^{}_{1}}$ at 
which the transition from SF state to PS or CDW state occurs. 
However, as demonstrated by the numerics below at
non-half-fillings, MFA fails to capture the supersolid phase.

\begin{figure}[t]
\includegraphics[angle=-90,width=0.48\linewidth]{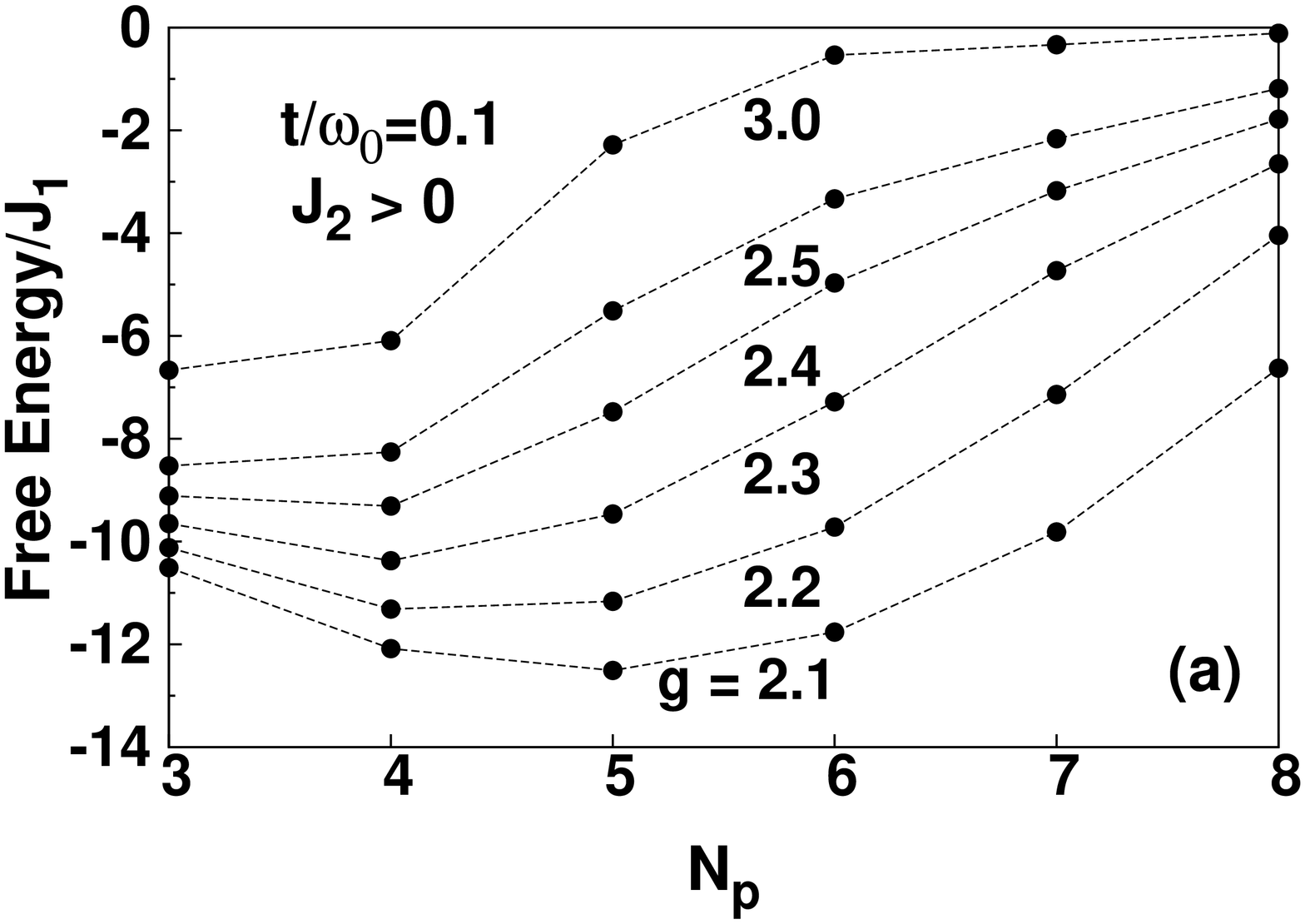}
\hspace{0.007\linewidth} \includegraphics[angle=-90,width=0.48\linewidth]
{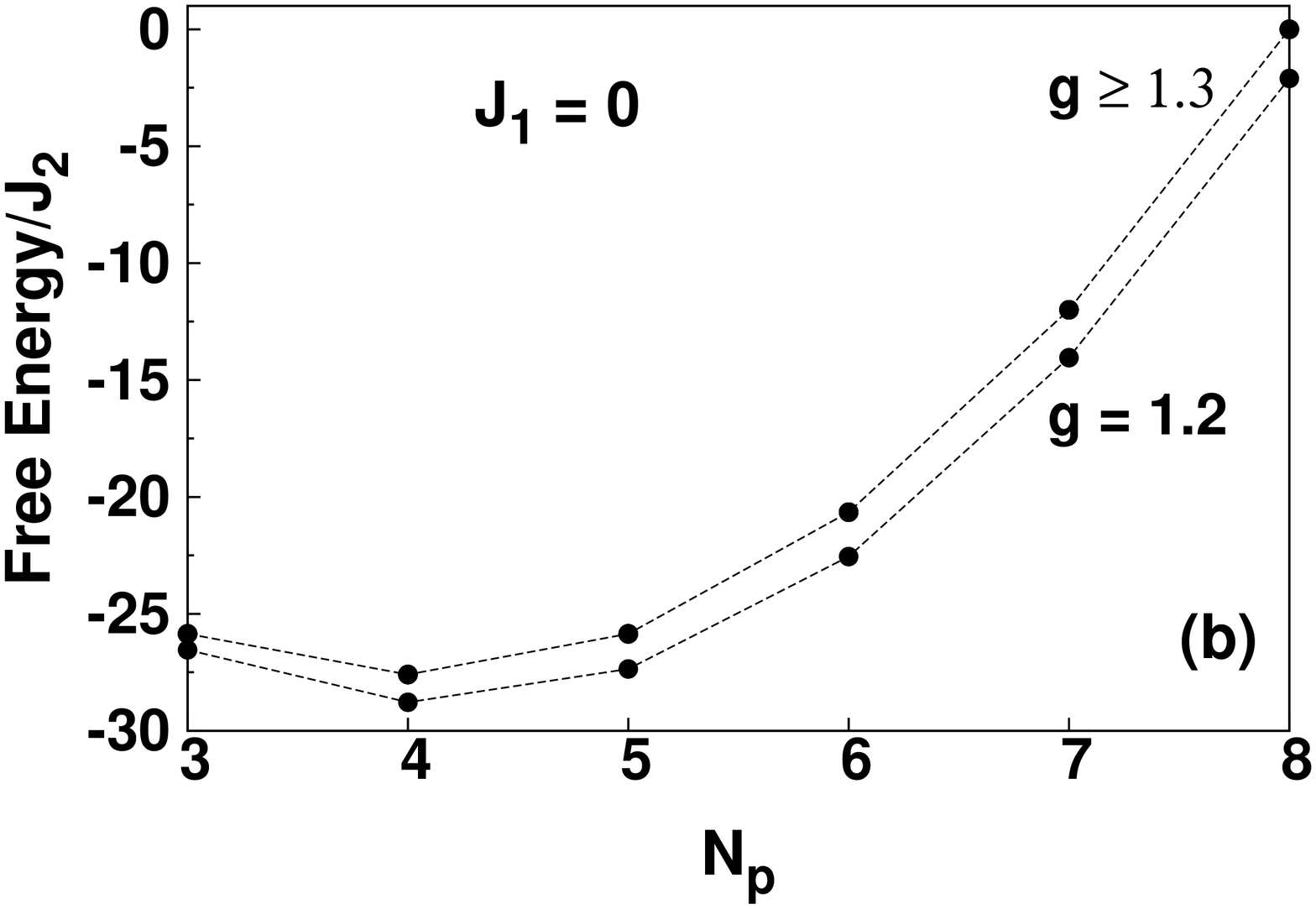}
\caption{Free energy at different fillings 
 and different values of $g$ for (a) $J_1 > 0$, $J_2 \neq 0$, and
$t/\omega_0 = 0.1$; and
(b) $J_1=0$, $J_2 \neq0$, and $t/\omega_0 =1.0$. }
\label{psep_ad0.1}
\end{figure}

\begin{figure}[b]
\includegraphics[angle=-90,width=0.95\linewidth]{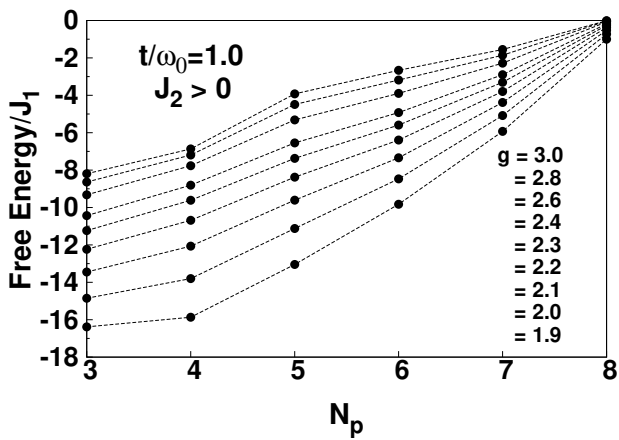}
\caption{Plot of free energy for different number of particles 
and various values of $g$ when $J_1 \neq 0 $, $J_2 \neq 0$,
and $t/\omega_0 = 1.0$.}
\label{psep_ad1.0.eps}
\end{figure}

We studied the stability of the phases by examining the nature of 
the free energy versus $N_p$ curves
at different b-p couplings $g$ (see Figs. \ref{psep_ad0.1} and
\ref{psep_ad1.0.eps}) and by using an analysis equivalent to
the Maxwell construction.
For the system at a given $N_p$, $g$, and $t/\omega_0$,
if the free energy point $P$ lies above the straight line joining the 
nearest two stable points $Q$ and $R$ (lying on either side of $P$)
 on the same free energy curve, then the system at $P$ breaks up into
two phases corresponding to points $Q$ and $R$.

For the range of parameters that we considered 
(i.e., $0.1 \le t/\omega_0 \le 1$ and $ g > 1 $), the
small parameter $t/(g \omega_0) < 1$.
The behavior of the system for $t/\omega_0 =0.1 $ is the same for both
 $J_2 = 0$ and $0 \neq J_2/J_1 $ [= $(J_1/\omega_0)  f_1(g)$] because $J_2/J_1$
is negligible in the latter case. Furthermore, for $J_2 =0$, the behavior
of the system as a function of $g$ is qualitatively the same for all values
 of $t/\omega_0$ as there is only one dimensionless parameter $J_z/J_1$
 involved in Eq. (\ref{heff}).

%*********************************************************************
%*********************************************************************
\begin{figure}[t]
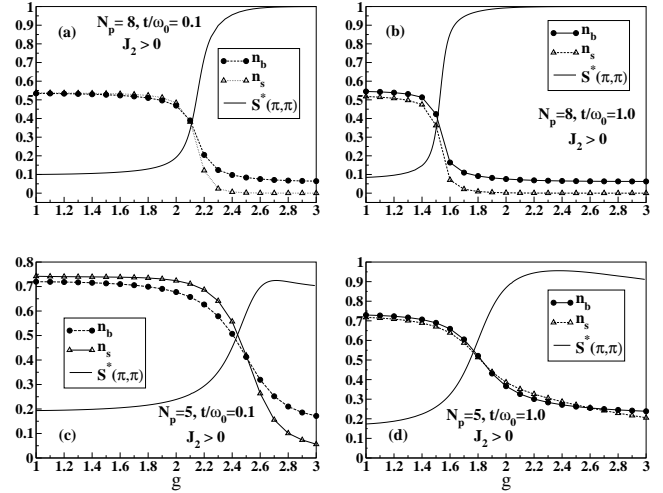

\includegraphics[angle=0,width=0.47\linewidth]{np8_bec.rho.sf_ad0.1.eps}
\hspace{0.025\linewidth}\includegraphics[angle=0,width=0.47\linewidth]{np8_bec.rho.sf_ad1.0.eps}\\
\vspace{0.028\linewidth}
\includegraphics[angle=0,width=0.47\linewidth]{np5_bec.rho.sf_ad0.1.eps}
\hspace{0.025\linewidth}\includegraphics[angle=0,width=0.47\linewidth]{np5_bec.rho.sf_ad1.0.eps}
\caption{Comparative plots of $S^{*}(\pi,\pi)$, $n_b$, and $n_s$ 
 when $J_1 \neq 0$, $J_2 \neq 0$, and
(a) $t/\omega_0 = 0.1$, $N_p=8$;  
(b) $t/\omega_0 = 1.0$, $N_p=8$;
(c) $t/\omega_0 = 0.1$, $N_p=5$;
 and (d) $t/\omega_0 = 1.0$, $N_p=5$.}
\label{np8_bec.rho.sf}
\end{figure}
%*********************************************************************

We will now analyze together, in one plot, the quantities
$n_b$, $n_s$, and the normalized structure factor
$S^{*}(\pi,\pi) = S(\pi,\pi)/S^{max}(\pi,\pi)$ where 
$S^{max}(\pi,\pi)$ corresponds to all particles occupying only one sub-lattice. 
At half-filling, we first observe that the system is
either a pure CDW or a pure SF.
For a half-filled system (i.e., $N_p =8 $) at $J_2 \neq 0$
and $t/\omega_0 = 0.1$ ($t/\omega_0 = 1.0$), 
we can see from Fig. \ref{np8_bec.rho.sf} (a) 
[Fig. \ref{np8_bec.rho.sf} (b)] that
the system undergoes a sharp (first order) transition to an
insulating CDW state at $g_c \approx 2.15$ ($g_c \approx 1.55$).
At $g=g_c$, while there is a sharp rise in
$S^{*}(\pi,\pi)$, there is also a concomitant sharp drop
in both the condensation fraction $n_b$ and the SF fraction $n_s$.
Furthermore, while $n_s$ actually goes to zero, $n_b$ remains finite
[as follows from Eq. (\ref{n_b})] at a value $1/N = 1/16$ which is
 an artifact of the finiteness of the system.
Larger values of $t/\omega_0$
for a half-filled system leads to lower values of $g_c $.
 This is in accordance with the MFA phase boundary
 Eq. (\ref{phaseboundary}) and the fact that 
$(J_z/J_1)\times (\omega_0/t)$ 
[$(J_2/J_1)\times (\omega_0/t)$] is monotonically increasing (decreasing)
function of $g$ for $g >1$. 

Away from half-filling, the system shows markedly 
different behavior compared to the half-filled situation.
For $N_p \le 4$ in the  $4 \times 4$ lattice considered here,
without actually presenting the details of the calculations,
we first note that there is no evidence
of a phase transition (for both $J_2 =0$ and $J_2 \neq 0 $).

In Fig. \ref{np8_bec.rho.sf} (c) drawn for $N_p = 5$,
although $S^{*}(\pi,\pi)$ displays a CDW transition at
a critical value $g_c =2.45$, $n_s$ does not go to zero
 even at large values of $g$ considered.
Furthermore, we see clearly from Fig. \ref{psep_ad0.1} (a) that, 
above this critical value of $g$, the  curvature of the free
energy curves suggests that the system at $N_p=5$ is 
an inhomogeneous mixture of CDW-state and SF-state.
 Thus away from half-filling, at small values of the adiabaticity
$t/\omega_0$, our HCB-system
 undergoes a transition from a SF-state to a PS-state at a critical 
b-p coupling strength [similar to the xxz-model in Fig. \ref{phase-diag2} (b)].

{\em However for $t/\omega_0$ not too small, when NNN hopping is present,
 the system shows a strikingly new 
 behavior for a certain region of the $g$-parameter space}.
 Let us consider the system at $N_p=5$, $t/\omega_0 = 1.0$, $J_1 \neq 0$,
and $J_2 \neq 0$. Fig. \ref{np8_bec.rho.sf} (d) shows that,
 above $g \approx 1.85$, the system enters a CDW state
 (as can be seen from the structure factor); however, it
continues to have a SF character as reflected by
the finite value of $n_s$.
Furthermore, Fig. \ref{psep_ad1.0.eps} reveals that the curve
is concave, i.e., the system is PS, only above $g = 2.0$.
This simultaneous 
presence of CDW and SF states,
without any inhomogeneity (for $1.85 < g < 2.1$),
implies that the system is a {\em  supersolid}. 
Similarly, for $6$ and $7$ particles as well,
we find that the system undergoes transition from a
SF-state to a 
SS-state
 and then to
a PS-state. This behavior is displayed
in Fig. \ref{phase-diag2} (a).

\begin{figure}[b]
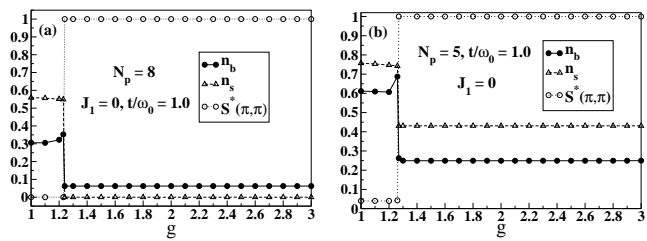

\includegraphics[angle=0,width=0.47\linewidth]{np8_nbns_j10_ad1.0.eps}
\hspace{0.025\linewidth}\includegraphics[angle=0,width=0.47\linewidth]{np5_nbns_j10_ad1.0.eps}
\caption{Comparison of $S^{*}(\pi,\pi)$, $n_b$, and $n_s$ 
when $t/\omega_0 = 1.0$, $J_2 \neq 0$ but
 $J_1 = 0$ and for (a) $N_p=8$; and (b) $N_p = 5$. 
}
\label{np8_nbns_j10_ad1.0.eps}
\end{figure} 
Finally, we shall present the interesting 
case of $J_1 =0$ and $J_2 \neq 0$
as a means of understanding the 
SS
phase
in the phase diagram of Fig. \ref{phase-diag2} (a). The physical scenario,
when $J_1$ can be negligibly small compared
to $J_2$, has been addressed in  Ref. [\onlinecite{sudhakar1}] for 
cooperative electron-phonon interaction in an one-dimensional system. 
 When $J_1 =0$, 
for large values of nearest-neighbor repulsion, it is quite natural that
 all the particles will occupy a single sub-lattice. 
However, the dramatic jump (at a critical value of $g$),
from an equal occupation of both sub-lattices
to a single sub-lattice occupation, is quite unexpected
(see Fig. \ref{np8_nbns_j10_ad1.0.eps}).
 For a half-filled system, above a critical point,
 all the particles get localized which results in an insulating state. 
This can be seen from Fig. \ref{np8_nbns_j10_ad1.0.eps} (a).
 One can see that (at $g \approx 1.23$)
 the structure factor dramatically jumps to its maximum value,
 while $n_s$ drops to zero and $n_b$ takes
 the limiting value of $1/16$ for 
reasons discussed earlier. This shows that above 
 $g \approx 1.23$, the system is in an insulating state 
with one sub-lattice being completely full. 
However, away from half-filling, the system
 conducts perfectly while occupying a single sub-lattice
because of the presence of holes in the sub-lattice.
For instance, from Fig. \ref{np8_nbns_j10_ad1.0.eps} (b)
drawn for $N_p=5$, we see that the structure 
factor jumps to its maximum value at 
$g \approx 1.26$, whereas $n_s$ drops
to a finite value which remains constant above $g = 1.26$.
We see from Fig. \ref{psep_ad0.1} (b), based on the curvature of the
free energy curves, that the 5-particle system does not phase
separate both above and below the CDW transition.
In fact, this single-phase-stability is true for any filling.
This means that the system, at any non-half filling and at $J_1 =0$,
exhibits supersolidity above a critical value of $g$!

{\bf Conclusions:}
We demonstrated that our BH model displays supersolidity in two-dimensions (2D).
Our effective Hamiltonian of Eq. (\ref{heff})
should be realizable in a 2D optical lattice.
Furthermore, our BH model too can be mimicked 
by designing HCB to move as extra particles in a 2D array of trapped molecules
with HCB coupled to the breathing mode of the trapped molecules \cite{zoller}.
In three-dimensions, supersolidity is more achievable
% in the following cases:  
(i) in general, due to an increase in the
ratio of NNN and NN coordination-numbers; and
(ii) in particular for bismuthates, because cooperative breathing mode enhances
the ratio of NNN and NN hoppings.
%*********************************************************************
%*********************************************************************
%*********************************************************************
%{\bf Acknowledgments}
%********************************************************************

S. Datta thanks Arnab Das for very useful 
discussions on implementation of Lanczos algorithm.
S. Yarlagadda thanks K. Sengupta, S. Sinha, A. V. Balatsky, R. J. Cava,
I. Mazin, and M. Randeria for valuable discussions.
This research was supported in part by the National Science Foundation
under Grant No. PHY05-51164 at KITP.
%********************************************************************

\end{document}